\documentclass[conference]{IEEEtran}
\usepackage[utf8]{inputenc}
\usepackage{tabu}
\usepackage{enumitem}
\usepackage[fleqn]{mathtools}
\usepackage{url}
\usepackage{svg}
\usepackage[hpos=0.7\paperwidth, 
            vpos=0.04\paperwidth,
            angle=90]{draftwatermark}

\title{Demystifying QUIC from the Specifications\vspace{-3mm}}
\author{
  \IEEEauthorblockN{
    Darius Saif\IEEEauthorrefmark{1}, Ashraf Matrawy\IEEEauthorrefmark{2}
  }
  \IEEEauthorblockA{
    Carleton University, Department of Systems and Computer Engineering\IEEEauthorrefmark{1}, School of Information Technology\IEEEauthorrefmark{2}\\
    Email: {dariussaif@cmail.carleton.ca, ashrafmatrawy@cunet.carleton.ca}\vspace{-3mm} 
  }
}
\begin{document}
\maketitle

\SetWatermarkText{Authors' Draft for Soliciting Feedback: \today}
\SetWatermarkColor[gray]{0.5}
\SetWatermarkFontSize{0.6cm}
\SetWatermarkAngle{0}
\SetWatermarkHorCenter{11cm}

\begin{abstract}
QUIC is an advanced transport layer protocol whose ubiquity on the Internet is now very apparent. Importantly, QUIC fuels the next generation of web browsing: HTTP/3. QUIC is a stateful and connection oriented protocol which offers similar features (and more) to the combination of TCP and TLS. There are several difficulties which readers may encounter when learning about QUIC:  i.) its rapid evolution (particularly, differentiation between the QUIC standard and the now deprecated Google QUIC), ii.) numerous RFCs whose organization, language, and detail may be challenging to the casual reader, and iii.) the nature of QUIC's cross-layer and privacy-centric implementation, making it impossible to understand or debug by looking at packets alone. For these reasons, the aim of this paper is to present QUIC in a complete yet approachable fashion, thereby demystifying the protocol from its specifications.
\end{abstract}

\begin{IEEEkeywords}
QUIC, HTTP/3, TLSv1.3, TCP
\end{IEEEkeywords}
\IEEEpeerreviewmaketitle

\section{Introduction}

The QUIC protocol was originally developed by Google around 2013 \cite{langley2017quic} and its motivations were to i.) improve the performance secure HTTP traffic, and ii.) address the role on-path middleboxes have on TCP traffic. QUIC is a stateful and connection-oriented transport protocol designed on top of UDP to provide many of the features that TCP and TLS traditionally would. These features include reliable data delivery, flow and congestion control, as well as authentication and encryption. QUIC implements advanced features like connection migration and \textbf{stream multiplexing}.

QUIC adopts a cross-layer protocol approach whereby both transport-layer (TCP) and application-layer features (TLS) are integrated. Unlike TCP and TLS for example, \textbf{connection establishment and key exchange occur concurrently by default}, which can save RoundTrip Times (RTTs). Additionally, QUIC provides the option to send data as independent streams within a single connection, which makes it more resilient to loss and Head-of-Line Blocking than TCP.

Google’s successful rollout of QUIC \cite{langley2017quic} led to it being proposed as a standard with the Internet Engineering Task Force (IETF) and the QUIC Working Group \cite{quicwg} was created in 2016. Google QUIC (gQUIC) and the IETF's QUIC coexisted for a period despite significant differences in their inner workings. By 2021, IETF QUIC was officially standardized under RFC 9000 \cite{rfc9000} and gQUIC was deprecated. Importantly, the \textit{de facto} transport protocol of HTTP/3 \cite{rfc9114} is QUIC.

Those looking to research or understand QUIC may face different challenges. First is QUIC's rapid evolution: older literature may reference gQUIC or draft standards of QUIC which can greatly vary with today's QUIC behaviors. Secondly, the target audience for the numerous RFCs \cite{rfc9000,rfc9001,rfc9002,rfc9221,rfc9368,rfc9369} detailing QUIC may not necessarily be casual readers. For example, RFC 9000 discusses \texttt{STREAM} frames before defining what frames are and through what means they are sent. Lastly, QUIC is a sophisticated protocol which combines multiple communication stack layers and its internal state machine cannot be inferred by simply observing traffic on the wire. The contributions of this paper are to: i.) present QUIC in a complete, yet approachable manner, ii.) provide a brief history on the motivation for QUIC and how it differs from TCP and TLS, iii.) highlight future work items of the QUIC Working Group.

\section{QUIC Packets \& Headers}
QUIC packets are transmitted over UDP datagrams. Multiple QUIC packets belonging to the same connection can be sent in a single UDP datagram, which is known as \textbf{coalescing packets}. Furthermore, QUIC offers both \textbf{packet and header protection}: packet protection encrypts payloads via TLSv1.3, then a portion of the encrypted payload is used to mask fields in the header for privacy. When packets are coalesced, their packet and header protections are applied separately.

QUIC packets use either a long or short header and have varying levels of packet and header protection. Packet payloads are made up of different \textbf{frame types} (outlined in Table 3 of RFC 9000). QUIC packet types, their cryptographic protections, and the header type they employ is presented in Table \ref{tab:quicPktTypes}. Note 0-RTT packets have no forward secrecy though.

{
\tabulinesep=1mm
\begin{table}[b]
  \centering
  \caption{QUIC Packet Types \& Header Variant}
  \begin{tabu} to 0.49\textwidth {|X[2.5]|X[2.6]|X[3.1]|}
     \hline
      \centering \textbf{Packet Type}&\centering \textbf{Header Type}&\centering \textbf{Protection Level}\\\hline      
      Version Negotiation&Special Long Header&No Protection\\\hline
      Initial&Long Header&Integrity\\\hline
      Handshake&Long Header&Confidentiality, Integrity\\\hline
      Retry&Long Header&Integrity\\\hline
      0-RTT&Long Header&Confidentiality, Integrity\\\hline
      1-RTT&Short Header&Confidentiality, Integrity\\\hline
 \end{tabu}
  \label{tab:quicPktTypes}
\end{table}
}

The long packet header is used during connection establishment, where TLSv1.3's cryptographic handshake takes place over \texttt{CRYPTO} frames. The long header is shown in Figure \ref{ch2:longHeader}. The Header Form bit is set to 1 for long headers and 0 for short headers. The Fixed Bit is set to 1, except for Version Negotiation packets. The Long Packet Type designates whether the packet is an Initial, 0-RTT, Handshake, or Retry packet. Depending on the packet type, the Type-Specific Bits are used for different cases. QUIC's version is also specified. The source and destination \textbf{connection identifiers} (IDs) and their lengths are discussed further in Section III.A. The payload of the packet then follows.

\begin{figure*}[t]
  \centering
  \includesvg[width=6in]{figures/longHeader.svg}
  \caption{Long Header Packet Format}
  \label{ch2:longHeader}
\end{figure*}

\begin{figure*}[t]
  \centering
  \includesvg[width=6in]{figures/shortHeader.svg}
  \caption{Short Header Packet Format}
  \label{ch2:shortHeader}
\end{figure*}

Once a connection is established, the short header is used within 1-RTT packets to reduce transmission overhead. The short header format is illustrated in Figure \ref{ch2:shortHeader}. The Header Form is set as 0 for the short header, and the Fixed Bit must be set to 1 -- otherwise, the packet will be ignored. The Spin Bit is used for passive latency monitoring. The Packet Number Length determines how many bytes the Packet Number field is. The Source Connection ID and its length are known by each endpoint and therefore do not need to be transmitted. The Key Phase field indicates which packet protection keys are used by the packet, supporting encryption key rotation.

Unlike TCP, which uses byte-level sequence and acknowledgment numbers, QUIC implements monotonically increasing \textbf{packet numbers}. QUIC packet numbers are unique within a connection and are not reused even for retransmissions -- this avoids TCP's retransmission ambiguity by clearly distinguishing which packets are being acknowledged and which were lost. This also simplifies QUIC's congestion control and offers more accurate RTT measurements, easier detection of spurious retransmissions, and a more straightforward Fast Retransmit mechanism. Packet numbers are divided into three separate spaces for: i.) Initial, ii.) Handshake, and iii.) 0-RTT and 1-RTT packets. This isolates each space from their varying levels of encryption and from incurring spurious retransmissions due to a different packet number space.

Many fields within QUIC's packets are implemented as \textbf{variable-length integers}, thus reducing their transmission size when values corresponding to such fields are small. This works by reserving the two most significant bits of the field to determine whether the field length is 1, 2, 4, or 8 bytes.

\section{Initiating \& Terminating QUIC Connections}

\begin{figure}[t]
  \centering
  \includegraphics[width=3.4in]{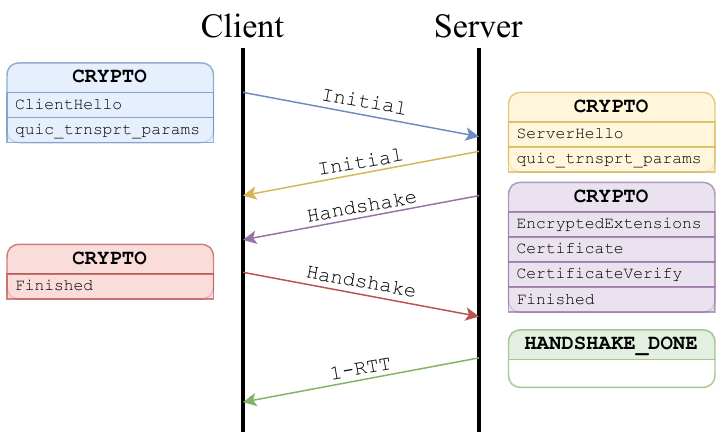}
  \caption{Simplified QUIC Handshake}
  \label{quicHandshake}
\end{figure}

As shown in Figure \ref{quicHandshake}, QUIC’s handshake starts with an Initial packet with a payload containing a \texttt{CRYPTO} frame housing TLSv1.3’s ClientHello. If early data is supported by the endpoints (and they previously shared a connection), 0-RTT packets may be coalesced with the Initial packet. The server responds with an Initial packet containing the ServerHello and immediately sends a Handshake packet to advance the key exchange procedure. Once an endpoint installs the 1-RTT keys and sends a Finish message in its Handshake packet, it begins sending 1-RTT packets and must stop sending 0-RTT packets.

QUIC's handshake includes a key exchange where the server is always authenticated, and the client is optionally authenticated. QUIC requires that clients maintain a buffer of at least 4096 bytes in case \texttt{CRYPTO} frames are received out of order \cite{rfc9000}.

Both endpoints signal their \textbf{\textit{quic\_transport\_parameters}} in their Initial packets \cite{rfc9001}. These parameters (listed in Table 6 of RFC 9000) are transmitted in an integrity-protected TLS extension and are authenticated after the QUIC handshake is completed. Endpoints are obliged to comply with all limits and restrictions set by their peer. Furthermore, the transport parameters advertised apply to all applications that an endpoint supports.

Some transport parameters from a previous connection to an endpoint must be stored and reused if a client wishes send 0-RTT packets to that server in a future connection. Furthermore, the TLSv1.3 \textit{early\_data} EncryptedExtension must be advertised by the endpoint in order to send early data with 0-RTT packets. It is advised that application (or other replayable) data not be sent in 0-RTT frames because of their susceptibility to replay attacks \cite{rfc9000}.

Connections can be terminated through various avenues: reaching the idle timeout threshold, receiving an immediate close, or by a stateless reset. The idle timeout threshold is the lower of the \textit{max\_idle\_timeout} advertised by either endpoint in their transport parameters. An immediate closure occurs upon receipt of a \texttt{CONNECTION\_CLOSE} frame. Two types of \texttt{CONNECTION\_CLOSE} frames exist, which signal either QUIC or application layer errors, respectively.

An endpoint may lose the state of a QUIC connection, in which case the third avenue of connection termination, stateless reset, may be used. Endpoints include an encrypted 16-byte stateless reset token for every connection ID they issue. When an endpoint receives packets that it cannot process, it will send a stateless reset. 

\subsection{Selection of Connection IDs}
QUIC defines these unique identifiers to recognize connections regardless of changes in the UDP or IP layers. Endpoints typically delegate multiple connection IDs for a given QUIC connection, and the receipt of any packet with a matching ID will be processed against that connection. Having multiple connection IDs facilitates connection migration and helps provide an additional layer of anonymity from onlookers, as connection IDs are not header protected. Connection IDs can be changed at any time during the connection.


Each endpoint selects the connection ID that its peer will use by encoding it into the Source Connection ID field of packets they send. The receiving endpoint must encode that ID into the Destination Connection ID of packets it responds with. If the client has not previously received Initial or Retry packets from the server, it will choose a random Destination Connection ID of at least 8 bytes to send in its very first Initial packet to the server. The Destination Connection ID's value is used towards determining the packet protection keys, which could change upon receipt of a server Retry packet. Connection IDs are also encoded in the QUIC transport parameters for data integrity verification.


New connection IDs are assigned to an endpoint through \texttt{NEW\_CONNECTION\_ID} frames. A client may choose to stop accepting packets associated to a connection ID by invalidating the ID through the \texttt{RETIRE\_CONNECTION\_ID} frame. Both \texttt{NEW\_CONNECTION\_ID} and \texttt{RETIRE\_CONNECTION\_ID} frames incorporate monotonically increasing sequence numbers.

\section{Authentication \& Validation}

\subsection{Address Validation}
Until an endpoint can verify that its peer is able to receive packets at the IP address it claims to own, the endpoint will limit the data it sends to three times the amount of data received from its peer. The purpose of this mechanism is to mitigate amplification attacks with spoofed addresses. Address validation is implicitly completed when an endpoint receives a valid Handshake packet from its peer, as this means that the peer successfully processed an Initial packet. This applies during both connection establishment \textit{and} connection migration.

QUIC also supports token-based address validation, whereby a server can validate the client address \textit{before} engaging in the cryptographic handshake. For this method of validation, the server must have sent the client a token (unique to that client) in either i.) a previous connection with a \texttt{NEW\_TOKEN} frame, or ii.) in a Retry packet. If a client has such a token, it must encode it in the Initial packet it sends to the server to initiate a new connection. Tokens can expire, therefore clients need only keep record of the newest token they receive from a server. Furthermore, servers construct these tokens in a way where they can distinguish whether they were communicated via a Retry packet or otherwise.

\subsection{Path Validation}
Path validation is necessary to ensure that both endpoints are still reachable during connection migration. This process confirms that packets sent to the peer's new path are received and that packets received by the migrating peer do not contain a spoofed source address.

An endpoint sends \texttt{PATH\_CHALLENGE} frames containing 8 bytes of unpredictable payload on the new path. The path is then validated once a \texttt{PATH\_RESPONSE} frame is received, relaying the same unpredictable payload from the challenge. Attempts to validate a new path may be abandoned if a response is not received within a given threshold.

\section{Connection Migration}
Connection migration is initiated when non-probing frames are sent by an endpoint from a new local address. Every QUIC frame listed Table 3 of RFC 9000 is non-probing, with the exceptions of \texttt{PATH\_CHALLENGE}, \texttt{PATH\_RESPONSE}, \texttt{NEW\_CONNECTION\_ID}, and \texttt{PADDING}.

Subsequently, if the peer accepts the migration, both address and path validation are performed. Upon successful validation, the endpoint resets its congestion controller state and any estimations about the peer's RTT. However, if the peer's UDP port changes not its IP address, it is not necessary to reset these values. Upon a failed validation, the endpoint must revert to communicating with the last validation address. If this is not possible, the connection will be closed.

\section{Version Negotiation}
There are two mechanisms of version negotiation, outlined by RFC 9368 \cite{rfc9368}: incompatible and compatible version negotiation. If a server does not support the version of QUIC selected by the client (and cannot decode the client's Initial packets), it will send a Version Negotiation packet, triggering \textbf{incompatible} version negotiation. 

As mentioned in Table \ref{tab:quicPktTypes}, the Version Negotiation packet uses a special long header, where the Fixed Bit, Long Packet-Type, and Type-Specific Bits are ignored by the client and the Version field must be set to 0. The packet payload contains a list of all supported QUIC versions of the server. The Version Negotiation packet is not cryptographically protected.

If the client has no mutually supported QUIC version with the list supplied by the server, it must abandon the connection attempt. If there is a mutually supported version, the client will send a new Initial packet, belonging to a new QUIC connection, encoding that version into the long header.

A client may attempt initiating a connection using a version of QUIC which the server does not support, but can partially understand. If the wire image and handshaking behavior of an unsupported version of QUIC are similar enough to a version that the server \textit{does} support, the server may be capable of parsing the Initial packet from its peer. In such a case, it is said that the server's version of QUIC is \textbf{compatible} with the client's. Compatibility is not symmetric; compatibility of one QUIC version to another does not guarantee that the opposite is true. RFC 9368 outlines that new versions of QUIC should explicitly define compatibility to other versions and vice versa.

If there is compatibility with the server's version of QUIC to the client's version, the server will be able to read the client's \textit{version\_information} transport parameter, which contains a list of QUIC versions the client supports. The server can then select a version from the list that it also supports and respond to the client in accordance with that version, encoding the version string in its long header. Compatible version negotiation is preferred, as it requires fewer round-trips.

\section{QUIC Streams}

\begin{figure}[t]
  \centering
  \includegraphics[width=3.3in]{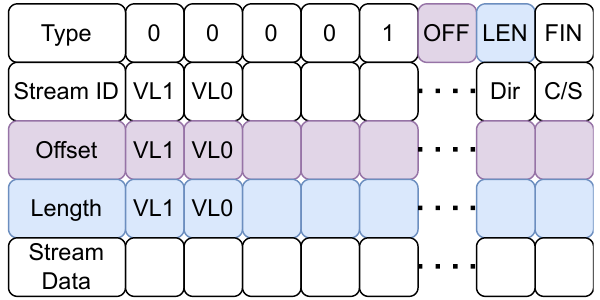}
  \caption{QUIC \texttt{STREAM} Frame Structure}
  \label{streamFrame}
\end{figure}

Streams reliably carry ordered byte-streams of data that can be concurrently interleaved with other streams within a QUIC connection. QUIC allows for either endpoint to initiate streams, and they can be either bi-directional or uni-directional. So long as \texttt{STREAM} frames belonging to different stream IDs are not coalesced, the loss of data on one stream will not affect any other stream, thus reducing HoLB. Furthermore, streams can be created (or ended) at any point during the lifetime of a QUIC connection. The structure of a \texttt{STREAM} frame is depicted in Figure \ref{streamFrame}.

The lowest three bits in the Type field signify important attributes for the remainder of the frame: they are the OFF, LEN, and FIN bits, respectively. When the OFF or LEN bits are set, they drive the presence of the Offset and Length fields. The FIN bit is set to mark the end of a stream. The Stream ID, Offset, and Length are encoded as variable-length integers, with their highest two bits as $VL_1$ and $VL_0$. 

The Offset field specifies the byte offset in the stream for the data contained in that particular frame, and, if not present, the offset is zero. It is important to note that while TCP conflates the sender's transmission order and receiver's delivery order, these are separated in QUIC through the use of packet numbers and the Offset field in \texttt{STREAM} frames.

The Length field specifies the number of bytes that the Stream Data field occupies. If zero, the Stream Data occupies the remainder of the QUIC packet. Streams are distinguished by the Stream ID field, which can be up to 62 bits long. Stream IDs cannot be reused within a QUIC connection. Table \ref{tab:streamConvention} shows the conventional meaning of the two least significant bits ($Dir$ and $C/S$) of a stream ID.

{
\tabulinesep=1mm
\begin{table}[b]
  \centering
  \caption{Stream ID Characteristic Conventions}
  \begin{tabu} to 0.49\textwidth {|X[2.4]|X[3.8]|}
     \hline
      \centering \textbf{Least Significant Bits}&\centering \textbf{Stream Type}\\\hline      
      00&Client Initiated, Bi-directional\\\hline
      01&Server Initiated, Bi-directional\\\hline
      10&Client Initiated, Uni-directional\\\hline
      11&Server Initiated, Uni-directional\\\hline
 \end{tabu}
  \label{tab:streamConvention}
\end{table}
}

Stream IDs of a given type must be used chronologically. Using client initiated bi-directional streams as an example, this means that the 0th and 4th stream IDs must have been exhausted before allocating stream ID 8.

Upon session establishment, endpoints negotiate the cumulative number of incoming streams that a peer is allowed to open via the \texttt{initial\_max\_streams\_bidi} and \texttt{initial\_max\_streams\_uni} transport parameters. During the connection, these values may be updated if an endpoint sends a \texttt{MAX\_STREAMS} frame. This allows peers to control the amount of concurrency they accept. Streams can remain open for the lifetime of a QUIC connection or can be terminated by the application. Additionally, QUIC supports application-driven relative prioritization of streams.

When an endpoint application no longer needs to read the data it is receiving on a stream, it can transmit a \texttt{STOP\_SENDING} frame for the other endpoint to cease data transmission. This prompts a \texttt{RESET\_STREAM} frame from the other endpoint, which terminates its sending portion of a stream.

\section{Flow Control}
The purpose of flow control is first to ensure that either peer is not receiving more data than it can process, and second to prevent the overcommitment of resources to malicious peers. QUIC's flow control is performed at both the \textbf{connection \textit{and} stream level}. For \texttt{CRYPTO} frames, flow control is handled independently to avoid excessive buffering. For application data, QUIC implements limit-based flow control at the individual stream level in addition to the overall connection. Initially, these values are advertised by each receiver in the transport parameters negotiated during connection establishment. During the course of the connection, these limits may be updated by signaling \texttt{MAX\_STREAM\_DATA} and \texttt{MAX\_DATA} frames, respectively. Only frames that increase the respective limits are processed; otherwise, they are ignored. When a sender has reached a limit set by the receiver, it will be unable to send further data until it is unblocked.

The frequency at which \texttt{MAX\_STREAM\_DATA} and \texttt{MAX\_DATA} frames presents a trade-off between higher resource overhead and potentially blocking a sender, depending on how often such frames are transmitted. A peer should strive to give out flow control credit greater than the bandwidth-delay product of the connection \cite{rfc9000}. It is also recommended that flow control updates be sent along with other frames, such as \texttt{ACKs}, to reduce overhead.

Streams (no matter how they are closed) must report a tally of the total bytes that were sent. This is so that the flow control algorithm of both endpoints can tabulate and agree on how much credit each stream has consumed. The \texttt{RESET\_STREAM} frame reports this value through the Final Size field. For \texttt{STREAM} frames with the FIN bit set, the final size of the stream is the sum of the Offset and Length fields.

\section{QUIC Datagrams}
RFC 9221 \cite{rfc9221} defines an additional avenue for transmitting application data over a QUIC connection. The optional \texttt{DATAGRAM} frame can only be used if both endpoints advertise their support in the QUIC transport parameters. These frames are subject to congestion control and are \texttt{ACK} eliciting, but QUIC itself makes no guarantees to retransmit lost \texttt{DATAGRAM} frames. Furthermore, \texttt{DATAGRAM} frames are not flow-controlled. Both streams and datagrams can be used simultaneously in a single QUIC connection and can be sent either in 0-RTT or 1-RTT packets.

\section{Congestion Control \& Loss Detection}
As RFC 9002 \cite{rfc9002} states, congestion control and RTT measurements are unified across packet number spaces and apply to the connection as a whole. Moreover, QUIC provides generic congestion control signals that can be used to support modern algorithms. Senders unilaterally choose which congestion control algorithm to use. Based on the algorithm, the transmission of packets should be paced or limited to prevent from bursts which could cause congestion or loss. QUIC packets which contain frames other than \texttt{ACK} or \texttt{CONNECTION\_CLOSE} count towards congestion control limits and are considered to be in-flight. Packets only containing \texttt{ACK} frames should not be paced. Packets that would cause the number of bytes in-flight to be larger than the congestion window must not be sent unless triggered by a Probe Timeout (PTO) expiry or when entering recovery. 

\subsection{Acknowledgment Behavior \& Generation}
QUIC's use of \texttt{ACK} frames play a key role in both congestion control and loss detection. The principles of TCP's Selective Acknowledgments (sACK) \cite{rfc2018} are present in QUIC with some notable differences. First, many more ranges are supported, which can be beneficial in high-loss network conditions. Second, QUIC does not implement ACK reneging, which is the practice of retracting a previously acknowledged chunk of data. The PTO mechanism is used to ensure that \texttt{ACK} frames are received by the peer.

Every packet that is received and processed by a QUIC endpoint is acknowledged. However, only ACK-eliciting frames trigger an \texttt{ACK} frame to be sent within the \textit{maximum\_ack\_delay} timer vowed under the QUIC transport parameters. Frames which are not ACK-eliciting (\texttt{PADDING}, \texttt{ACK}, and \texttt{CONNECTION\_CLOSE}) are acknowledged when the next necessary \texttt{ACK} frame is sent.

\texttt{ACK} frames consists of a Largest Acknowledged field (the largest packet number the peer is acknowledging), an ACK Delay field (used towards refining RTT measurements), and one or more ACK Ranges. Acknowledged packet numbers belong to the same packet space as the \texttt{ACK} frame itself. Furthermore, acknowledgments are irrevocable.

The number of ranges an \texttt{ACK} frame contains is controlled by its ACK Range Count field. ACK ranges themselves consist of Gap and ACK Range Length fields. The First ACK Range is an integer value representing contiguous packet numbers being acknowledged (from the Largest Acknowledged downward). The Gap field is used to indicate contiguous packet numbers which are not acknowledged (starting one packet number less the lowest packet number in the preceding ACK Range).

The receiver can limit the number of ACK Ranges it remembers and sends, but including acknowledgments for older packets can help reduce spurious retransmissions in case a previous \texttt{ACK} frame was lost. \texttt{ACK} frames should be able to fit in a single QUIC packet, and, if not, older ranges may be omitted.

\subsection{RTT Measurement}
QUIC uses RTT measurements and peer-reported delays to build a statistical representation of the network path. Three values are calculated to meet this end: the minimum over a window of time \textit{min\_rtt}, an exponentially weighted moving average \textit{smoothed\_rtt}, and the mean deviation \textit{rttVar}. RTT samples (\textit{latest\_rtt}) are taken from \texttt{ACK} frames which newly acknowledged at least one ACK-eliciting packet. The sample is generated by taking the time delta between the sending time of the packet number matching the \texttt{ACK} frame's Largest Acknowledged field and the time the \texttt{ACK} was received. The \texttt{ACK} frames may also report an ACK Delay field, which specifies the time period that the peer intentionally waited before sending the \texttt{ACK}, but this is only factored in when computing \textit{smoothed\_rtt} and \textit{rttVar}.

\subsection{Acknowledgment-Based Loss Detection}
Unlike RTT measurements and congestion control, loss detection operates separately for each packet number space. Packets are deemed lost when i.) the packet has not been acknowledged and was sent prior to an acknowledged packet, and ii.) the packet has hit QUIC's time threshold or \textit{kPacketThreshold}. RFC 9002 recommends an initial \textit{kPacketThreshold} value of 3 packets \cite{rfc9002}. The time threshold is derived from the larger of the \textit{smoothed\_rtt} and \textit{latest\_rtt} measurements (with a floor of at least 1 millisecond). These thresholds provide some tolerance before declaring loss in case packets have simply arrived out-of-order.

An ACK-eliciting probe packet must be sent when an ACK-eliciting packet has not been received by an endpoint in the time frame it expects, as per the PTO mechanism. A second probe packet may be sent to avoid consecutive PTO expirations or to transmit data from multiple packet number spaces. This helps QUIC recover from the loss of tail packets or acknowledgments. An expired PTO timer does not indicate that packet loss has occurred. The PTO timer is reset once an \texttt{ACK} frame is received and acknowledges new packet numbers. The PTO value is the summation of the \textit{smoothed\_rtt}, \textit{max\_ack\_delay} transport parameter, and four times the \textit{rttVar} (with a floor of at least 1 millisecond).


When QUIC packets are lost, the necessary frames are retransmitted in new packets with higher packet numbers. Packet numbers are never reused in QUIC.

\section{UDP Datagram Requirements \& Discovery}
QUIC requires that: i.) UDP datagrams cannot be fragmented at the IP layer, and ii.) a network path can support a maximum datagram size (that is, the largest UDP payload transmitted over a single datagram) of at least 1200 bytes.

Fragmentation adds the complexity of reassembly which also degrades performance, especially over networks with packet loss and reordering. These restrictions imposed by QUIC ensure that TLS handshake data can be transmitted without fragmentation and allow QUIC to discover network paths that support datagram sizes larger than 1200 bytes. This is accomplished with either Path Maximum Transmission Unit Discovery (PMTUD) \cite{rfc1191,rfc8201} or Datagram Packetization Layer PMTUD (DPLPMTUD) \cite{rfc8899} -- both of which QUIC natively supports.

\section{QUIC Version 2}
QUIC version 2 is defined under RFC 9369 \cite{rfc9369} and versions 1 and 2 are mutually compatible. QUIC version 2 is not necessarily meant to replace version 1. Primarily, it seeks to mitigate ossification by middleboxes that may begin to assume that the bytes contained in the Version field of QUIC's long headers will always be \texttt{0x00000001} or that the Initial packet's key derivation formula will be version-invariant. This could lead to middleboxes not properly processing new versions of QUIC. Version 2 of QUIC \textbf{does not introduce any new functionality}. Rather, it introduces several semantic and trivial changes:
\begin{itemize}
    \item The Version field in long headers is delegated the value \texttt{0x6b3343cf}
    \item TLS keys and nonces and the salt used for Initial key derivation are updated
    \item The meanings of the 2-bit Long Packet Type field are changed
\end{itemize}

\section{Future Work Items for QUIC}
The QUIC-WG references several projects which are in the draft phase and, at the time of writing, not yet RFC standards. Such projects include:
\begin{itemize}
    \item Multipath support over a single QUIC connection by introducing various new frame types and a path identifier
    \item qlog: An event-based logging schema to aid in debugging QUIC, because of its encrypted nature 
    \item The \texttt{RESET\_STREAM\_AT} frame, to allow for resetting a stream while guaranteeing delivery of stream data to a certain byte offset
    \item \texttt{ACK\_FREQUENCY} and \texttt{IMMEDIATE\_ACK} frames for endpoints to request changes in their peer's acknowledgment behavior
    \item Generating Routable QUIC connection IDs so that load balancers can correctly route QUIC packets with addresses that have been migrated
\end{itemize}

\bibliographystyle{IEEEtran}
\bibliography{IEEEabrv,references.bib}

\end{document}